\documentstyle[a4,11pt,epsf,graphicx]{article}
\begin{document}
\setlength{\baselineskip}{8pt}  {\flushleft {\scriptsize {\rm
Lecture notes:} \\ {\sc The Second Tah Poe School on Cosmology:
Modern Cosmology (TPCosmo II)} \\ {\rm 17-25 April 2003, Naresuan
University, Phitsanulok, Thailand}}} \\  \\
\renewcommand{\baselinestretch}{0.7}
\setlength{\baselineskip}{12pt} {\flushleft{\LARGE {\bf
Introductory Overview of Modern Cosmology}}}
\\
\\
\begin{quote}{\large {\rm {\bf Burin Gumjudpai}}}$^{\dag}$
\\ \setlength{\baselineskip}{12pt}
{\small{\rm Institute of Cosmology and Gravitation}} \\
{\small{\rm University of Portsmouth}}\\ {\small{\rm Portsmouth
PO1 2EG, United Kingdom}} \\ \\ {\footnotesize $^{\dag}$e-mail:
{\sl burin.gumjudpai@port.ac.uk}} \\ \\
\end{quote}
\renewcommand{\baselinestretch}{0.7}
\setlength{\baselineskip}{14pt} \begin{quote} {\bf Abstract:}\\ A
basic modern picture of the universe is given here. The lectures
start from the historical ideas of a static universe. Then I move
on to Newtonian cosmology and derive the main cosmological
equations in the framework of Newtonian mechanics for the sake of
simplicity. With a qualitative description of general relativity,
the expansion of the universe and elementary idea of the hot Big
Bang models are introduced. The problems of the hot Big Bang,
inflation and structure formation are explained here in simple
language.
\end{quote}

\newpage
\setlength{\baselineskip}{12pt}
\renewcommand{\baselinestretch}{0.5}
\tableofcontents \setlength{\baselineskip}{15pt}

\renewcommand{\baselinestretch}{1.2}
\setlength{\baselineskip}{15pt}
\section{Introduction}
{\sl What are we in? How big is it? How did it begin?}
\\{\sl How will it end?}. . .
\\
\\
Some of the oldest questions in the world are surely those that
ignited one of the world's oldest subjects. The above questions
are as such responsible for cosmology. Answers and explanations
for these questions were sought from many concepts, cultures and
beliefs. From ancient times until today our civilization has been
gradually adding many small pieces of the jigsaw, either on
mythological, philosophical or scientific aspects of our new
picture of the universe. Different viewpoints of our universe vary
from culture to culture. In science, cosmology started when we
believed that our physical laws applied on earth can also be
applied in any other parts of the universe. This is part of the
thought that we are not special and we do not occupy a special
place in the universe. Physics that works on earth must be the
same physics that works everywhere in the universe.

\section{The Newtonian Universe}

It was not until the 1930s that Hubble began to realise that our
Milky Way is not the only one main structure in the universe; in
fact our Milky Way is just one of many galaxies. Modern cosmology
was born. Historically Newton, without knowing that the sun is
just one of billions of stars of the Milky Way, and without having
any ideas about what lies beyond the Milky Way, attempted to apply
his three laws of motion and his law of gravitation to work in all
locations in the universe.
\\

Newton realized that the universe can not have finite size since
it would collapse quickly to the centre of mass. In the infinite
universe there is no centre of the mass sphere and this fact
allows the universe to be static agreeing with the beliefs of the
people in the 17th century. At smaller scales gravity can bring
about the formation of stars and smaller celestial objects. His
belief in a static universe was expressed in the second edition of
the Principia:
\\
\begin{center}
{\it The fixed stars, being equally spread out in all points of
heavens, cancel out their mutual pulls by opposite attractions.}
\end{center}

\subsection{Olbers' paradox and the Newton's static and infinite universe}
\subsubsection{The paradox}

\begin{figure}[t]
\begin{center}
\includegraphics[width=8cm,height=8cm,angle=0]{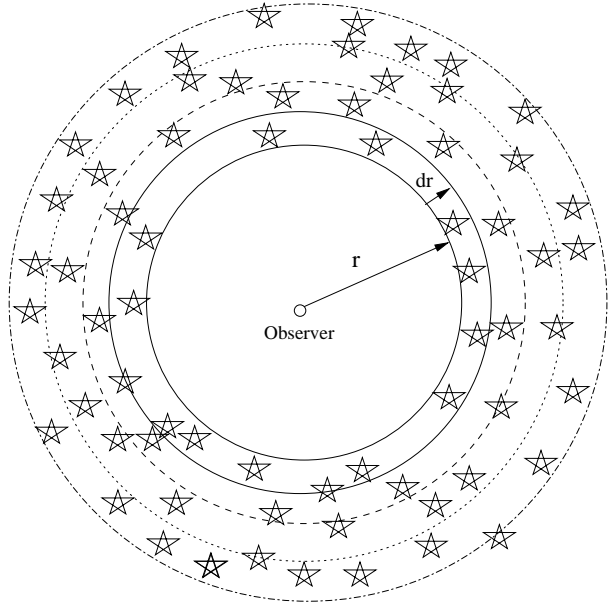}\label{fig1}
\end{center}
\caption{Observer's sphere with infinite sky}
\end{figure}

Newton's static infinite universe has a problem called {\bf
Olbers' paradox} (Digges 1576 and Olbers 1826). The paradox is
simply: {\it why is the night sky dark?} This paradox can be seen
in figure 1. In the figure we are the observers living at the
centre of the imaginary sphere. $r$ is the radial distance from
us. Each shell with $dr$ thickness occupies volume $4\pi r^2 dr$.
The luminosity of stars $L$ is proportional to the number of stars
$N$ and the number of stars is proportional to the volume of the
shell. The flux light density is therefore
\begin{equation}
Flux \:=\: \frac{L}{4\pi r^2}\:\propto\:  \int \frac{4\pi r^2
dr}{4 \pi r^2 } \:\propto\: \int dr
\end{equation}

If the universe is infinite, $\int dr \rightarrow \infty$ and the
flux of night-sky's light should be infinite.  The other problem
arose when the {\bf Copernican principle}, which roughly says that
we are not occupying a special place in the universe, was applied
to this model  (Halley 1721). This principle will be introduced
formally later. The universe is isotropic about any point in the
universe then the attraction force from stars in one side in
figure 1 must be equal to the attraction force from stars in the
opposite side. If there are infinite number of shells, the force
from each side becomes stronger and stronger but the total force
needs to be canceled out to zero. The universe is in balance with
very high instability. This situation needs a very high degree of
isotropy, otherwise this would unbalance the forces and pull the
matter toward one side, leading to collapse! Local forces that
govern local motions of planets would disturb this low stability.
\\

The problem of the Newtonian infinite universe is in fact what is
called in mathematics a Dirichlet problem of a potential theory.
To determine force, we need to know the boundary conditions. In a
universe with a uniform, finite and bounded distribution of
matter, we can determine the gravitational force at each point by
boundary conditions. In Newton's infinite, unbounded (edgeless)
and centreless universe, gravitational potential diverges at
infinite distance and the gravitational field depends on boundary
conditions at infinity. Another problem is the high instability as
mentioned earlier. To keep the Newtonian infinite universe static,
we need to set one point in the universe to possess zero
gravitational field but non-zero elsewhere. It does not sound
reasonable. Then the problems can not be solved without modifying
the dynamical law itself.

\subsubsection{The way out of the paradox}

Before the invention of General Relativity (GR) in 1915, gravity
was thought to act simultaneously (action at a distance). This
means matter particles can feel the force from other particles
instantaneously without time elapsing. In GR, light has finite
speed and so does the gravitational interaction. If we just add
the idea of gravity propagating at the speed of light, in the
finite age of the universe (approximately 10 billion years), we
don't feel force and don't see any light from distances beyond 10
billion light years. Olbers' paradox hence is solved since the
light from beyond 10 billion light years has not reached us yet.
Moreover, later when I introduce the expanding universe in which
the galaxies are moving away from each other, creating a redshift
in light and making distance between us and galaxies further and
further away, we will have an even better resolution for  Olbers'
paradox. More detailed discussion can be found in Refs.
\cite{Harrison} and \cite{burin}.

\subsection{Cosmological principles}

Modern cosmology relies on the basic important assumption called
{\it Cosmological principle} which is a more general version of
the {\bf Copernican principle}. The Copernican principle states
that the Earth is not the centre of the universe i.e. we are not
living at a special location in the universe and the universe must
be homogenous. {\bf Homogeneity} of the universe means that the
universe has the same property at any regions from point to point.
The cosmological principle includes the Copernican principle
together with {\bf isotropy} of the universe. Isotropy of the
universe means that the universe looks the same from all
directions. In Figure 2, the vector field is homogenous but
obviously it is not isotropic since when we observe it from a
fixed point in the picture, it looks different from different
directions. Figure 3 shows isotropy about one point but it fails
to include homogeneity. These two properties are therefore
separated and we need both of them to complete the cosmological
principle. Mathematically, homogeneity and isotropy are
respectively invariant properties under translation and rotation
transformations.
\\

 We indeed do know that at small scales the universe is not homogenous and not isotropic otherwise any
structures e.g. galaxies, stars, planets and humans would not even
exist. However provided that we consider the universe on average
on large scales, it looks approximately homogenous and isotropic.
Strong evidence for this is the cosmic microwave background (CMB)
observed in 1992 by the COBE mission to be very smooth to at least
one part in $10^5$ \cite{bennett}.

\begin{figure}[t]
\begin{center}
\includegraphics[width=6cm,height=6cm,angle=0]{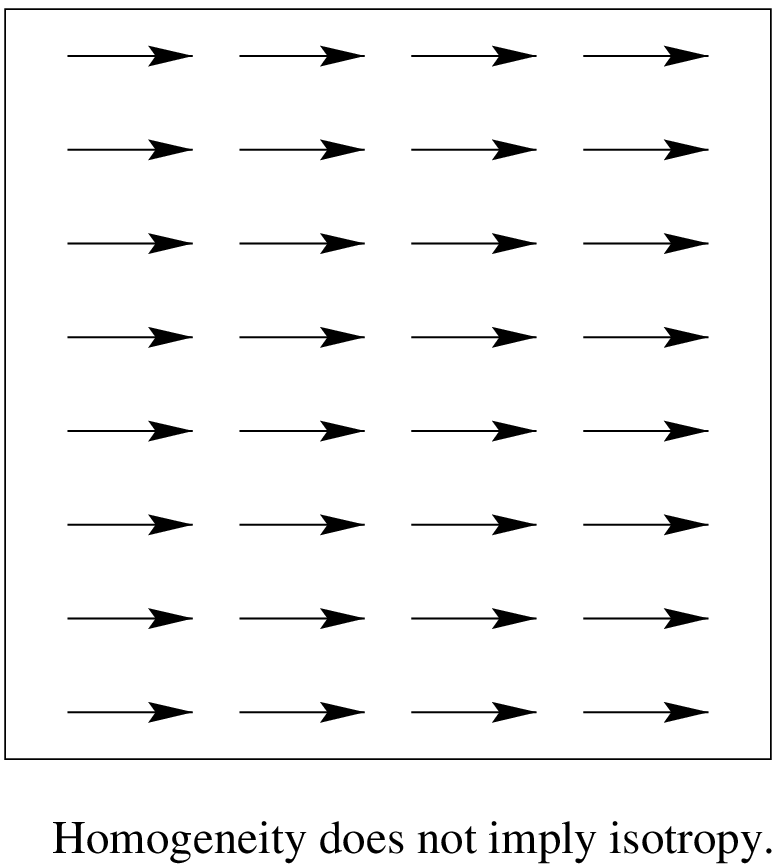}
\end{center}
\label{fig2} \caption{Homogenous vector field fails to be
isotropy}
\end{figure}

\begin{figure}[t]
\begin{center}
\includegraphics[width=6cm,height=6cm,angle=0]{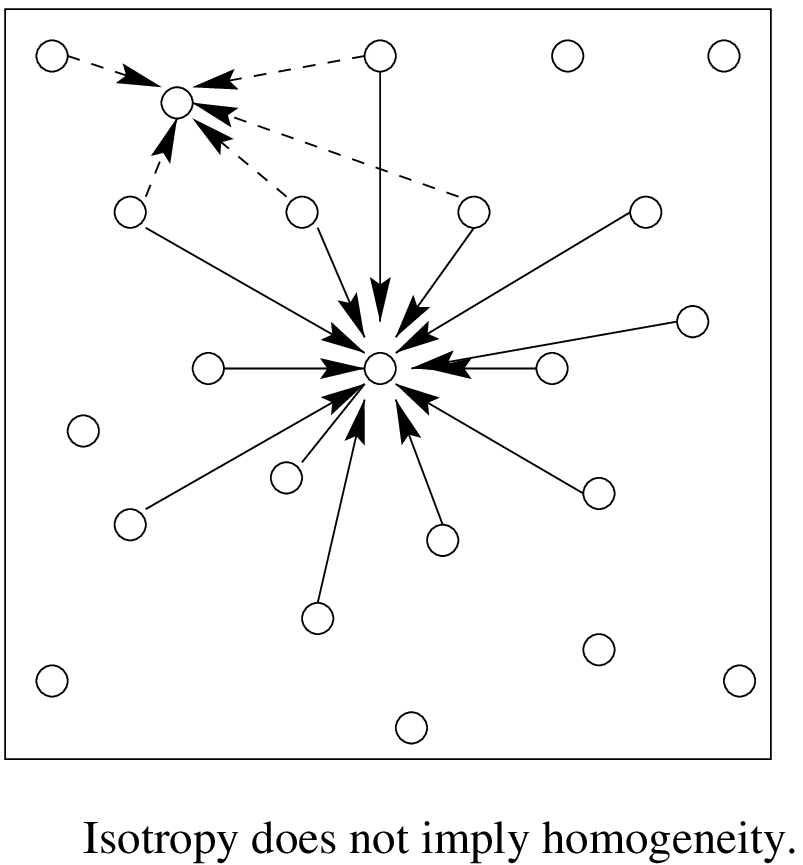}
\end{center}
\label{fig3} \caption{Isotropy does not imply homogeneity}
\end{figure}

\section{The Expanding Universe}
\subsection{The non-static universe}   Before 1915
cosmologists believed that the cosmos is static. From a Newtonian
viewpoint, in addition the universe is also infinite. The
solutions of GR in 1915 suggested that the universe should not be
static. Einstein therefore added {\bf cosmological constant}
$\Lambda$ into his field equation in order to obtain a static
solution and this was later admitted by him to be his greatest
blunder. Many relativists (de Sitter 1917, Friedmann 1922
\cite{fr}, Lema{\^{i}}tre 1927) found solutions (either infinite
or finite) in which universe is either expanding or collapsing. In
1929 the expansion of the universe was discovered by Hubble
(\cite{Hub},\cite{Hub1} and \cite{Hub2}). Later in 1934 Milne and
McCrea discovered that the Friedman-Lema{\^{i}}tre type model can
be recovered by using Newtonian Mechanics. In 1936, Robertson and
Walker gave the general form of the metric for a homogeneous and
isotropic universe, called the
Friedmann-Lema{\^{i}}tre-Robertson-Walker (FLRW) metric.
\\

We will take advantage of Milne and McCrea's work \cite{mc} by
using the Newtonian treatment to obtain the main cosmological
equations. We later discuss very briefly about GR. For a nice
historical discussion, I refer readers to Refs. \cite{Harrison}
and \cite{aegean1}.

\subsection{Hubble's law}

During the 1920s and 1930s, cosmologist began to realise that the
distant nebulae are not only stardust but in fact other galaxies
outside our Milky Way. The universe is much bigger than people at
that time thought. In 1929, Hubble found that the universe does
not stay static but is expanding \cite{Hub}. This was noticed by
observing that these distant galaxies' spectra are redshifted and
they are even more redshifted at further distance. Hubble found
the empirical law
\begin{equation}
{\mathbf{v}} \:=\:H_{0}{\mathbf{R}}  \label{Hubble}
\end{equation}
which was later dubbed after his name, {\bf Hubble's law}. This
law implies that the further galaxy is, the faster it moves away
from us. Here $H_{0}$ is the {\bf Hubble constant} at the time
$t_0$. The Hubble constant is in fact the proportionality constant
of ${\mathbf{v}}$ and ${\mathbf{R}}$. The law is not exactly true
since at the smaller scale the universe is neither completely
homogenous nor isotropic and there is also the peculiar velocities
from local gravitational forces. The law also breaks down at large
distances, when it is no longer a good approximation.
\\

\begin{figure}[t]
\begin{center}
\includegraphics[width=13cm,height=12cm,angle=0]{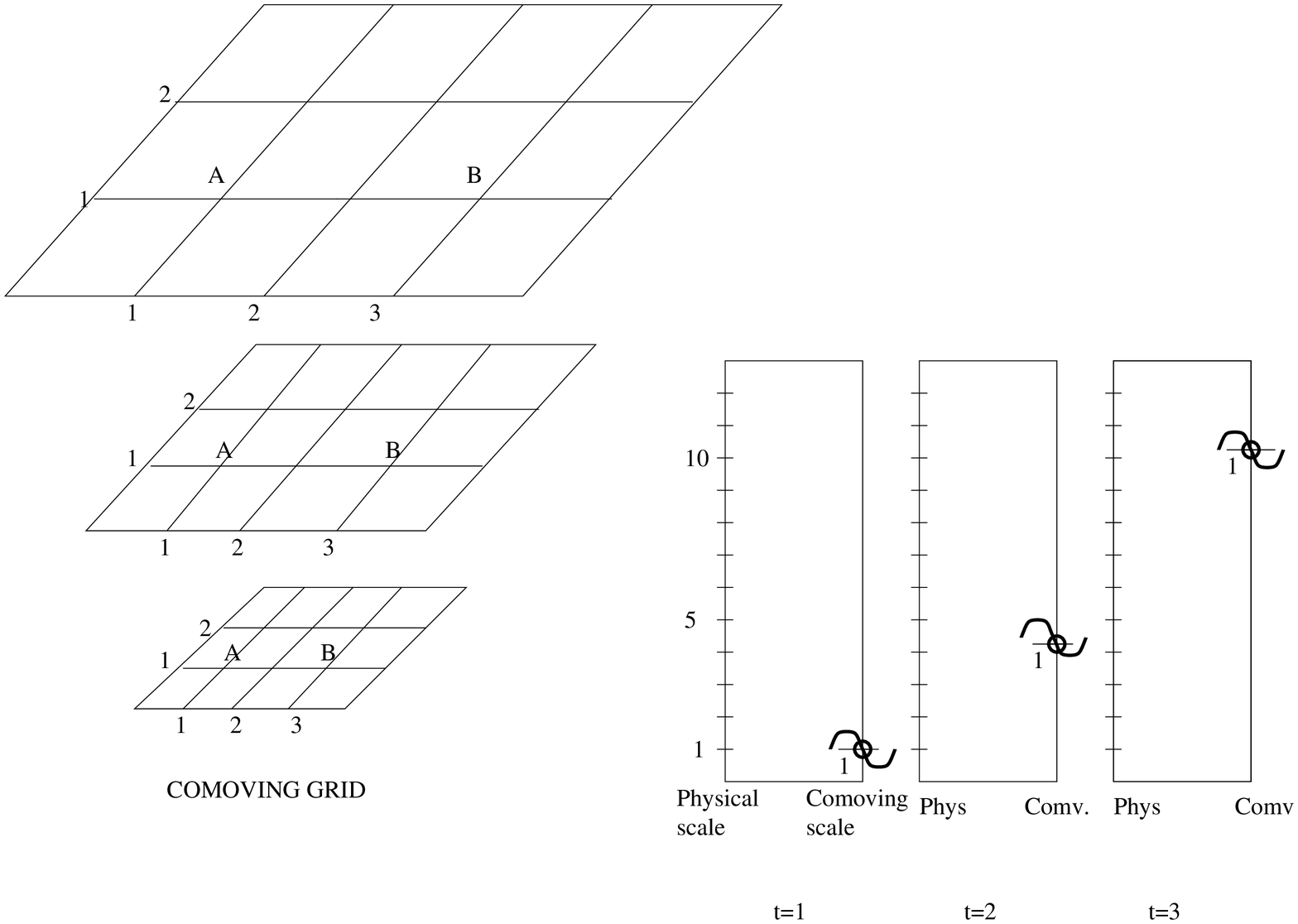}
\end{center}
\label{fig4} \caption{Expanding comoving grids and comoving galaxy
on a physical ruler}
\end{figure}

The expansion of the universe leads us to the concept of expansion
of space. Space {\it itself} is expanding and we need to introduce
the so-called {\bf comoving coordinates} $\mathbf{x}$. These
coordinates are moving along with the expansion of space. The {\bf
physical distance} $\mathbf{R}$ is given by
\begin{equation}
{\mathbf{R}} \:=\: a(t){\mathbf{x}}
\end{equation}
  where $a(t)$ is the {\bf scale factor}. Here we assume
homogeneity and isotropy of space. Figure 4 shows the expanding
coordinates in one and two dimensions. The recession velocity of
galaxies ${\mathbf{v}}$ is in the same direction as the
displacement from us ${\mathbf{R}}(t)$ and is given by
\begin{eqnarray}
{\mathbf{v}} \:&=&\: \dot{\mathbf{R}}  \nonumber \\
               &=&\: |\dot{\mathbf{R}}| \frac{\mathbf{R}}{|{\mathbf{R}}|}
\end{eqnarray}
 where dot denotes differentiation with respect to time e.g. $\dot{R} =
\frac{dR(t)}{dt}$. Using equation (\ref{Hubble}) , then
\begin{eqnarray}
{\mathbf{v}} \:&=&\: \frac{\dot{a}}{a}{\mathbf{R}} \nonumber \\
               &=&\: H {\mathbf{R}}
\end{eqnarray}
where
\begin{equation}
\fbox{ \parbox{2.3cm} {\[ H = \frac{\dot{a}}{a} \]} }
\end{equation}
 is the {\bf Hubble parameter} which is time-dependent. The constant $H_{0}$ usually means $H$ at
 the present age of the universe.
There is a subtle question about why are not we expanding? Why
does not the distance between the sun and the earth increase? The
answer is that the expansion only has a dominant effect on large
scales (inter-galactic scale). At scales smaller than this scale
we hardly see any effect of expansion since it is dominated by
local gravity.
\\

Another doubt is something to do with Special Relativity (SR). As
we see the distant galaxy moving away from us very fast, can it go
away faster than the speed of light? Yes, it can. This expansion
does not violate special relativity. Since special relativity
governs the speed of objects that are just passing one another.
That speed can not exceed the speed of light. Here in our
circumstance, it is the distant cosmic expansion of space, not two
objects passing each other.

\subsection{Redshift}

Redshift is the result of Doppler's effect in physics. The
expansion of the universe stretches the photon's wavelength. Light
spectra of elements which should be observed in some range of
$\lambda$ are shifted to the red end of the spectrum. The redshift
$z$ is defined by
\begin{equation}
z \:\equiv \:\frac{\lambda_{ob}-\lambda_{emit}}{\lambda_{emit}}\:
=\: \frac{\lambda_{ob}}{\lambda_{emit}} - 1 \:=\: \frac{a_{ob}
x}{a_{emit} x} - 1 \:=\: \frac{a_{ob}}{a_{emit}} - 1 \nonumber
\end{equation}
hence
\begin{equation}
z+1 = \frac{a_{ob}}{a_{emit}}
\end{equation}
The subscript $ob$ and $emit$ denote the observed and emitted
value of the wavelength.

\subsection{Energy and matter species} Matter and energy are convertible according to $E=mc^2$.
Matter or energy density in the universe can be classified into
three species:
\begin{itemize}
\item dust
\item radiation
\item cosmological constant ($\Lambda$) and scalar fields ($\phi$).
\end{itemize}

Dust is the term for non-relativistic matter e.g. particles with
small speed compared to the speed of light. Other species is
radiation which is relativistic particles moving with the speed
closed to that of light. Dust usually means baryons, electrons
etc. and radiation here means photons, neutrinos. The cosmological
constant is another form of energy that generates repulsive force
to prevent the universe from collapsing due to gravity. This
repulsive force can accelerate the expansion of the universe. We
call types of energy that can accelerate the universe {\bf dark
energy}. Scalar fields are candidates for dark energy and they can
be considered as a time-varying cosmological constant.

\subsection{Newtonian gravity}

In  Newtonian mechanics
\begin{equation}
{\mathbf{F}} \:=\: \frac{GM}{r^2}m \hat{{\mathbf{r}}}
\end{equation}
where $\mathbf{F}$, $M$, $m$, $G$ denotes force, mass, test mass
and the Newtonian gravitational constant. $r$ is the distance
between these two masses and $\hat{r}$ is a unit vector pointing
from $m$ to $M$. The force is always attractive and
$\frac{GM}{r^2} \hat{{\mathbf{r}}}$ is the gravitational
acceleration $\mathbf{g}$. The potential energy is given by
\begin{equation}
P.E. \:=\: -\frac{GMm}{r}
\end{equation}
The potential function is
\begin{equation}
\Phi \:=\: -\frac{GM}{r}
\end{equation}
For spherically symmetric and uniformly distributed mass (see
figure 5, similar to what happens in electrostatics, the potential
is constant within the sphere due to the symmetry property.
Particles in this sphere feel no force according to
\begin{equation}
{\mathbf{g}} \:=\: -\nabla\Phi
\end{equation}
The mass outside the sphere feels a force as if all the mass of
the sphere is located at the centre, and this force is independent
of the radius of the sphere, $R$.
\begin{figure}[t]
\begin{center}
\includegraphics[width=7cm,height=7cm,angle=0]{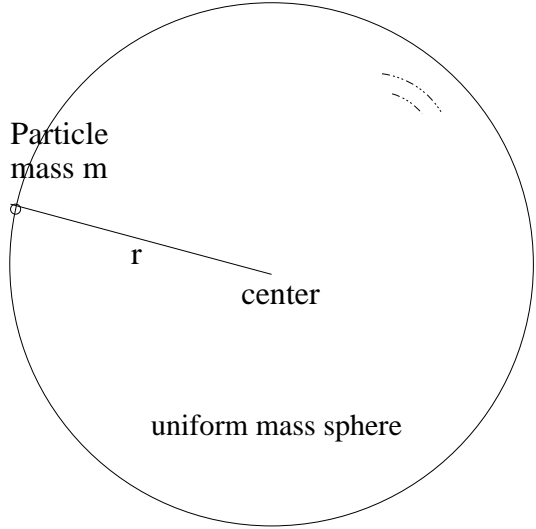}
\end{center}
\label{fig5} \caption{Sphere of uniformly distributed mass}
\end{figure}
\\

In the picture used by Milne and McCrea the sphere of mass is
expanding with a velocity $v = \dot{R}(t)$. The spherical mass
stays homogenous and isotropic (as viewed from centre of the
sphere) during the expansion.
\\

Let us look at the energy conservation law of a particle of mass
$m$ at the surface of this sphere. Here the total mass of our
sphere is $ \rho V $ where $\rho$, $V$ are the mass density and
the volume. $V = \frac{4}{3}\pi R^3$. Then the total energy of
this particle is
\begin{eqnarray}
U & = & \frac{1}{2}m \dot{R}^2\, + \, P.E. \nonumber \\
  & = & \frac{1}{2}m \dot{R}^2\, - \, \frac{4}{3}G \rho \pi R^2 m \nonumber
\end{eqnarray}
Multiplying both sides by $ 2/(mR^2)$,
\begin{equation}
\bigg(\frac{{\dot{R}}}{R} \bigg)^2 \, =  \, \frac{8 \pi G}{3}\rho
\, + \, \frac{2U}{mR^2} \label{fm1}
\end{equation}

This equation is just another form of the energy conservation law.

\section{Cosmological Equations}

\subsection{Friedmann equation} Now we will include the idea of
expansion in the energy conservation by putting $R(t) = a(t)x$
into equation (\ref{fm1}). Hence
\begin{equation}
\bigg(  \frac{\dot{a}}{a} \bigg)^2  \, = \, \frac{8 \pi G}{3}\rho
\, + \, \frac{2U/{mx^2}}{a^2} \label{fku}
\end{equation}

The fraction $\frac{2U}{mx^2}$ is a constant. When space expands,
 the $x$ value remains the same since it {\it comoves} with the
space. Total energy $U$ also remains constant. Let us define
$kc^2$ as $-2U/(mx^2)$ where $c$ is the speed of light. $k$ is the
{\bf curvature of space} which describes geometry of the universe.
If we add Einstein's cosmological constant $\Lambda/3$ to the
equation, it will be equivalent to adding a constant amount of
energy to the universe. This does not violate energy conservation.
Now with $H = \dot{a}/a $ the equation becomes
\begin{equation}
\fbox{\parbox{6cm}  {\[ H^2  \,= \, \frac{8 \pi G}{3}\rho \, - \,
\frac{kc^2}{a^2} + \frac{\Lambda}{3}  \]}}   \label{friedmann}
\end{equation}
This equation is known as the {\bf Friedmann  equation}. According
to the Copernican principle the equation we have here can be
applied to any regions of the universe. The Friedmann equation is
crucial in cosmology. It relates and constrains how the scale
factor $a$ evolves given the total density $\rho$ (amount of
energy and matter and hence geometry), $k$ and $\Lambda$ of the
universe.

\subsection{Fluid equation}

 We model the  matter and
radiation of the universe as a perfect fluid (fluid has no
viscosity and no heat conduction in comoving coordinate). We shall
start from look the first law of thermodynamics
\begin{eqnarray}
dU  \,& =& \, TdS + dW \nonumber  \\
    \,& =& \, TdS - pdV  \label{U}
\end{eqnarray}
where $S,T,W,p$ are entropy, temperature, work done and pressure
respectively. This law is also an energy conservation law. Here we
assume that the equation of state is of the form
\begin{equation}
p\,=\, p(\rho) \label{state}
\end{equation}
That is to say $p$ is an explicit function of $\rho$ only, and we
assume that there are no other external forces. For dust $p = 0$
and radiation $p= {\rho c^2} /3 $, but for $\Lambda$ or dark
energy, $p < 0$ (negative pressure). $dS$ vanishes since
reversible (adiabatic) expansion is assumed here. $dV$ is
$d(\frac{4}{3}\pi R^3) = 4 \pi R^2 dR$. Equation (\ref{U}) finally
becomes
\begin{equation}
dU \,= -p 4 \pi R^2 dR \label{U2}
\end{equation}
The total relativistic energy of particles in a sphere is
\begin{eqnarray}
U \,& =& \, Mc^2  \nonumber \\ \,&=&\, \rho V c^2 \nonumber \\
 dU  \,& =& \, 4 \pi R^2 \rho c^2 dR + \frac{4}{3} \pi R^3 c^2
 d\rho \label{U3}
\end{eqnarray}
If we differentiate equation (\ref{U2}) with respect to $t$ and
then combine it with equation (\ref{U3}), we get
\begin{equation}
 -p 4 \pi R^2 \dot{R} \,=\,4 \pi R^2 \rho c^2 \dot{R} + \frac{4}{3} \pi R^3 c^2
 \dot{\rho}
\end{equation}
Rearranging terms in this equation and writing it in terms of the
scale factor $a(t)$, we finally obtain the {\bf fluid equation}
\begin{equation}
\fbox{\parbox{5cm} {\[ \dot{\rho} + 3 \frac{\dot{a}}{a} \Big( \rho
+ \frac{p}{c^2} \Big) \:= \: 0 \]}}
\end{equation}
The equation of state (equation (\ref{state})) can be rewritten as
\begin{equation}
p \,=\, \rho c^{2}w
\end{equation}
where $w=0$ and $1/3$ for dust and radiation, while $w<0$ for
$\Lambda$ or dark energy. Fluid equation is then simply
\begin{equation}
\fbox{\parbox{5cm} {\[   \dot{\rho} + 3H \rho \big( 1 + w \big)
\:= \: 0 \]}} \label{flu}
\end{equation}

Fluid equation, likes Friedmann equation, is in fact energy
conservation law. The first term $\dot{\rho}$ tells us how fast
density changes (e.g. dilutes) and the second term is the lost of
kinetic energy from fluid into gravitational potential energy.

\subsection{Acceleration equation} The acceleration equation tells us
how rate of expansion of the universe changes i.e. slowing down or
speeding up. The equation is in fact a mixture of Friedmann and
fluid equations and these two equations are in fact energy
conservation law in mechanics and thermodynamics respectively.
After differentiating the Friedmann equation with respect to time
and using the fluid equation, we finally obtain
\begin{equation}
\fbox {\parbox{5cm} {\[  \frac{\ddot{a}}{a} \:=\: -\frac{4\pi
G}{3}\rho \big( 1 + \frac{3w}{c^2}\big) + \frac{\Lambda}{3} \]} }
\label{acc}
\end{equation}
The good feature of the acceleration equation is that it does not
contain $k$ and we can use this equation regardless of the
geometry of the universe. From the equation it seems that universe
is decelerating. When we neglect the small value of the
cosmological constant and the universe is dominated by $w<-1/3$
fluid (dark energy) with $p < -\rho/3$, it could make $\ddot{a}$
positive and will accelerate the universe. Indeed the recent
observation from Type Ia Supernovae strongly supports that the
universe now is accelerating \cite{perl}! From the viewpoint of
high energy physics we can have dark energy in the form of scalar
field that can yield negative pressure and hence accelerate the
universe. These scalar fields are in general called {\bf
Quintessence}, the name of the fifth element in ancient Greek. For
a good start to quintessence, Ref. \cite{sth} is recommended here.

\subsection{Solutions of equations}

\subsubsection{Solution of the fluid equation}

For simplicity, from now on we will work in the units where $c
\equiv 1$. The equation of state of the cosmological perfect fluid
is then just $p = w \rho$. The fluid equation (\ref{flu}) with the
help of basic calculus can be rearranged as
\begin{equation}
\frac{d}{dt}\Big(\rho a^{3(1+w)}\Big) \,=\, 0
\end{equation}
Obviously we see that $\rho a^{3(1+w)}=$ constant, and therefore
we get
\begin{equation}
\rho \propto a^{-3(1+w)}
\end{equation}
An alternative way is to notice that equation (\ref{flu}) is
separable differential equation
\begin{eqnarray}
\frac{1}{\rho} \frac{d \rho}{dt} &=& -3 \frac{\dot{a}}{a} (1+w)
\nonumber
\\
\int \frac{1}{\rho} d \rho &=&  -3(1+w) \int \frac{1}{a} da
\nonumber
\\
\ln \rho &=& -3(1+w) \ln a  + C \nonumber \\ \rho &=& e^{C}
a^{-3(1+w)} \nonumber
\end{eqnarray}
\begin{equation}
\fbox{\parbox{5cm}{\[ \rho =
\rho_{0}\bigg(\frac{a}{a_0}\bigg)^{-3(1+w)} \label{solflu} \]}}
\end{equation}
where $e^C = \rho_{0} a_{0}^{3(1+w)}$. Here we have assumed that
$w$ has constant value. The subscript $0$ denotes the value at
present\footnote{For simplicity in many textbooks, authors usually
re-scale the present size of scale factor to one ($a_{0} \equiv
1$).}. Here I introduce the {\bf e-folding number},
\begin{equation}
 N(t) = \ln \frac{a(t_{after})}{a(t_{initial})} \label{N}
\end{equation}
for use in later sections. The e-folding number tells us the
amount of expansion in log scale and is very useful when we
consider inflationary expansion of the universe. Suppose that we
consider the amount of expansion from the present era when the
scale factor is $a_{0}$ to sometime $t$ in the future. Then $N(t)
= \ln \big[a(t)/a_0 \big]$ and
\begin{equation}
a(t) = a_0 e^{N(t)}
\end{equation}
Using the e-folding number in the fluid equation (\ref{flu}), we
have
\begin{equation}
\dot{a} \frac{d\rho}{da} = -3 \frac{\dot{a}}{a} \rho (1+w)
\nonumber
\end{equation}
and
\begin{equation}
a \frac{d\rho}{da} = \frac{d \rho}{{[\frac{a}{a_{0}}]}^{-1}
d(a/a_{0})} = \frac{d \rho}{d \ln {(\frac{a}{a_{0}})}} = \frac{d
\rho}{dN} \nonumber
\end{equation}
Therefore
\begin{equation}
\frac{d\rho}{dN} = -3 \rho (1+w)
\end{equation}
and equation (\ref{solflu}) can be written as
\begin{equation}
\rho = \rho_{0} e^{-3(1+w)N}
\end{equation}

\subsubsection{Solution of the Friedmann equation} The exact
solution of the Friedmann equation can be found easily when we
assume that the universe is flat\footnote{Inflation theory of
universe predicts that universe has {\it flat geometry} ($k=0$).
Moreover CMB detectors such as BOOMERanG, MAXIMA
\cite{boom,Maxima} and Wilkinson Microwave Anisotropy Probe (WMAP)
\cite{wmap1} have already released datasets confirming that the
universe is very close to flat.} and $\Lambda$ has negligible
value. Using equation (\ref{solflu}), the Friedmann equation reads
\begin{equation}
\bigg(\frac{\dot{a}}{a} \bigg)^2  = \frac{8 \pi G}{3} \rho_{0}
 \bigg(\frac{a}{a_0}\bigg)^{-3(1+w)} \nonumber
\end{equation}
Then the solution is
\begin{equation}
 \fbox{\parbox{4cm}{\[ a \, = \, a_{0} \bigg(\frac{t}{t_0}\bigg)^{2/[3(1+w)]} \]}}  \label{scale}
\end{equation}

where $t_0$ is arbitrary constant. As the universe evolves, one
particular component dominates the universe at some period of
time. Following equation (\ref{scale}), domination of each type of
fluid leads to a particular type of the expansion kinematics. For
example in a universe with $w < -1/3$, dark energy fluid dominates
according to equation (\ref{acc}), and the expansion of the
universe is therefore accelerating.

\subsubsection{Evolution of density with time}
Putting together solutions (\ref{solflu}) and (\ref{scale}) we
obtain
\begin{equation}
 \fbox{\parbox{3.5cm} {\[ \rho(t) = \rho_{0} \bigg( \frac{t}{t_0} \bigg)^{-2} \]} }
\end{equation}
regardless of the value of $w$ (regardless of fluid types).

\subsubsection{Evolution of Hubble parameter with time}

Differentiating equation (\ref{scale}) with respect to time and
dividing the result again by $a$, the Hubble parameter evolves
with time as
\begin{equation}
 \fbox{\parbox{4.2cm} {\[ H(t) = \bigg[ \frac{2}{3(1+w)}\bigg]
 \frac{1}{t} \label{hubl} \]} }
\end{equation}
The Hubble parameter at the present time is called the Hubble
constant, $H_0$, which is given in terms of the dimensionless
constant $h$ as
\begin{center} $ H_0 = 100 h $ km/s/Mpc
\end{center}
The expansion increases $100 h$ km/s for every distance increase
of $1$ Mpc. (The distance $1$ pc $\simeq 3.261$ light years
$\simeq 3.086 \times 10^{16}$ m.) The value of the constant $h$ is
between $0.55$ and $0.75$. WMAP data (February 2003) leads to $h =
0.71^{+0.04}_{-0.03}$ \cite{wmaph}.

\section{Simple Toy Models}
\subsection{The dust-filled universe}
Dust (non-relativistic particles) is pressureless ($p=0$ or
$w=0$). In the dust-dominated universe solutions (\ref{solflu})
and (\ref{scale}) become
\begin{eqnarray}
\rho(t) &=& \rho_{0} \bigg( \frac{a(t)}{a_0}\bigg)^{-3}
\label{rhodust}
\\ a(t) &=&  a_{0} \bigg(\frac{t}{t_0}\bigg)^{2/3}  \label{adust}
\end{eqnarray}
respectively.  The equation tells us that the density decreases
with the expanding volume of the universe. In the dust universe,
 $H$ evolves as
\begin{equation}
 H(t) \: = \: \frac{2}{3t}
\end{equation}
implying that the universe will stop expansion ($H=0$) when $t
\rightarrow \infty$. Let us assume that dust always dominates the
universe. With $h = 0.71$ then we can find that
\begin{center}
$ H_0 \simeq 2.30 \times 10^{-18}$ s$^{-1}$
\end{center}
At the beginning of expansion $t=0$ s, so that the age of the
dust-filled universe today is
\begin{center} $t_{0} = \frac{2}{3 H_0} \simeq 2.90 \times  10^{17}$ s
\end{center}
or $9.20 \times 10^9$ years! Dust has been dominating the universe
longer than other type of fluids therefore this value is an
approximate age of our universe.

\subsection{The radiation-filled universe}
Radiation (relativistic particles) has $p=\rho/3$ or $w=1/3$.
Carrying out the same procedure as we performed in the dust case,
the solutions (\ref{solflu}) and (\ref{scale}) become
\begin{eqnarray}
\rho(t) &=& \rho_{0} \bigg( \frac{a(t)}{a_0}\bigg)^{-4}
\label{rhorad} \\ a(t) &=& a_{0} \bigg(\frac{t}{t_0}\bigg)^{1/2}
\label{arad}
\end{eqnarray}
respectively.  The radiation-filled universe expands slower than
the dust-filled universe. Density in both cases decreases with
$t^2$ but the radiation density decreases with $a^4$  not $a^3$ as
in dust case. The extra $a$ comes from the redshift effect on
relativistic particle's wavelength. We can obtain the relationship
between redshift and scale factor from the Hubble law. At
extra-galactic scale, for a small distance apart $dr$ the
recession velocity differs by $dv$. According to the Hubble law,
\begin{equation}
dv = H dr = \frac{\dot{a}}{a} dr  \label{dv}
\end{equation}
$H$ is assumed constant within the variation of small distance
$dr$. The Doppler effect stretches the radiation's wavelength
$\lambda$ by $d\lambda = \lambda_{ob} - \lambda_{emit}$. By
Doppler's law
\begin{equation}
\frac{d\lambda}{\lambda_{emit}} \:=\: \frac{dv}{c}  \label{dlam}
\end{equation}
The time lapses in light traveling a distance $dr$ is $dt=dr/c$.
Inserting equation (\ref{dv}) into (\ref{dlam}) we get
\begin{equation}
\frac{d\lambda}{\lambda_{emit}} \:=\: \frac{(\dot{a}/a)dr}{c}
\:=\:\frac{1}{a}\frac{da}{dt}dt \:=\: \frac{da}{a}
\end{equation}
We can see that $\lambda \propto a$ therefore frequency $\nu
\propto a^{-1}$. The energy density of radiation ($\rho_{rad}$) is
\begin{equation} \epsilon = n h \nu = n h \frac{c}{\lambda}
\propto \frac{1}{a}
\end{equation}
where the quantum of energy is $E = h\nu$. Here $h$ is Planck's
constant and $n$ denotes the photon number density. Therefore
redshift can add a factor $a^{-1}$ to the decreasing of
$\rho_{rad}$, and therefore yields $\rho \propto a^{-4}$ instead
of $ a^{-3}$.

\subsection{The dust and radiation-filled universe}
In the dust-radiation universe we have contributions from both
types of fluid. The Friedmann equation is now
\begin{equation}
\bigg(  \frac{\dot{a}}{a} \bigg)^2  \, = \, \frac{8 \pi G}{3} \Big
[\rho_{dust} + \rho_{rad} \Big]
\end{equation}
where $\rho_{dust}$ and $\rho_{rad}$ are given by (\ref{rhodust})
and (\ref{rhorad}) respectively. To get the exact solution for
this mixed fluid equation is not easy. In the universe containing
two components of fluid, dust falls off slower than radiation,
hence after sometime from the beginning, there must be a time when
$\rho_{rad} \:=\:\rho_{dust}$, and it is called {\bf
matter-radiation equality}. After that, dust will start to
dominate the universe. At early times, if we assume a Big Bang, it
should be radiation that dominates the universe, and we can
approximately use the solutions of the radiation-filled case.
After the equality we then can approximately use all equations of
the dust-dominated case. When dust becomes dominant, expansion
becomes faster (compare equations (\ref{arad}) and (\ref{adust})).
Figure 6 illustrates schematically the evolution of dust and
radiation density in the dust-radiation mixed universe. We can see
from figure 6 that after dust becomes dominant at equality, both
radiation and dust density fall off faster than before. This is
because the scale factor in the dust case increases faster than in
the case of radiation.

\begin{figure}[t]
\begin{center}
\includegraphics[width=10cm,height=10cm,angle=0]{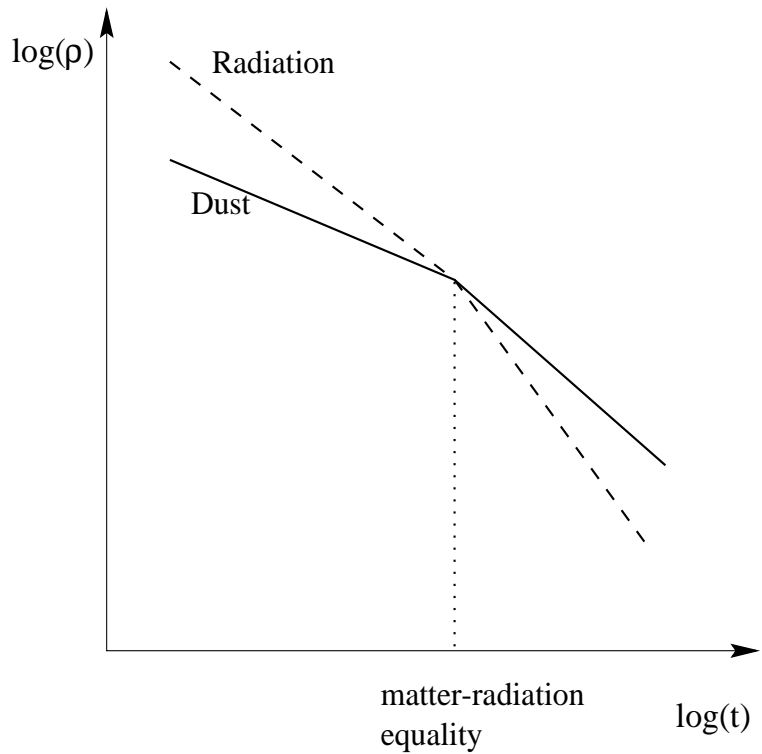}
\end{center}
\label{fig6} \caption{Schematic diagram of the evolution of
density in the dust-radiation dominated universe (picture from
Ref. \cite{liddle})}
\end{figure}

\subsection{Spatial curvature and fate of the universe}
The spatial curvature $k$ can play an important role in the
evolution of the universe. We have already looked in detail at the
$k = 0$ case. In the situation that $k \neq 0$, the term $k/a^2$
will dominate Friedmann equation quickly since it falls off much
slower than both $\rho_{dust}$ and $\rho_{rad}$.\\
\begin{itemize}
\item For $k < 0$ after $k/a^2$ becomes dominant, the Friedmann
equation reads
\begin{equation}
\bigg( \frac{\dot{a}}{a}\bigg)^2 \:=\: -\frac{k}{a^2}
\end{equation}
The solution of this equation is just
\begin{equation}
a \propto t
\end{equation}
yielding that the universe will expand forever!\\

\item For $k > 0$, the expansion will slow down because $-\frac{k}{a^2}$ reduces the Hubble rate.
Now we can not neglect $\rho$ since then $H^2 < 0$. Eventually the
curvature term balance the matter term
\begin{equation}
\frac{8 \pi G}{3} \rho =  \frac{k}{a^2}
\end{equation}
so that the expansion stops
\begin{equation} H \;=\;0
\end{equation}
From now the spatial curvature term becomes dominant and the
universe starts to collapse. Notice that if we substitute $-t$ for
$t$, the Friedmann equation remains unchanged \cite{liddle}. This
means that the equation is time-reversible and the universe will
evolve reversely to where it first starts expanding. We can see
the schematic evolution curves in figure 7.
\end{itemize}

\begin{figure}[t]
\begin{center}
\includegraphics[width=12cm,height=9cm,angle=0]{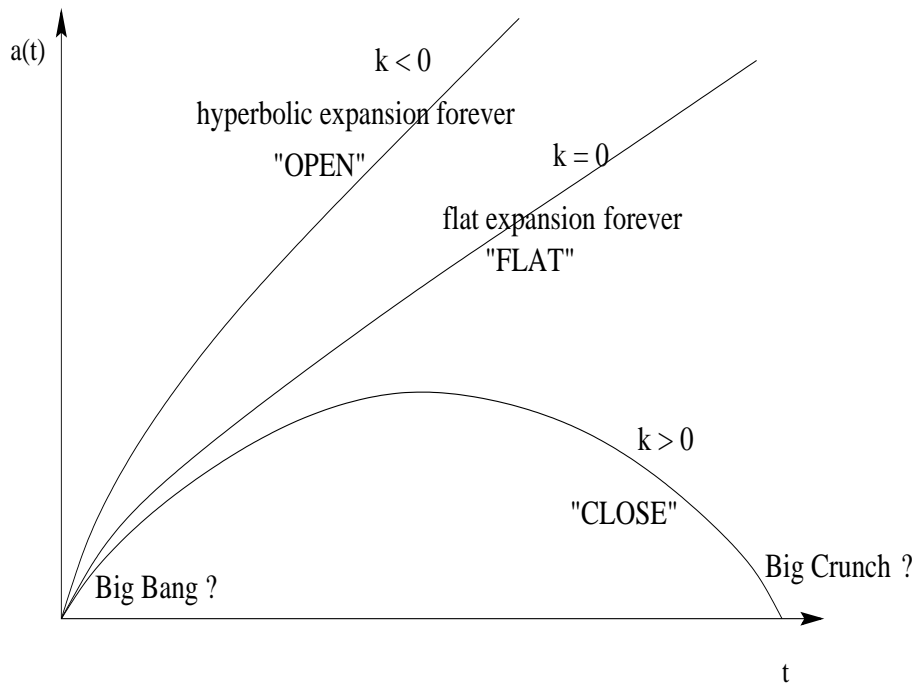}
\end{center}
\label{fig7} \caption{Curvature and fate of the universe (expand
forever or re-collapse)}
\end{figure}

\section{The General Relativistic Universe and the Big Bang}

      The story of the hot Big Bang arose from revolution of physics
ideas and astronomical observations in the early 20th century.
After Einstein, in 1915, discovered his General Theory of
Relativity (GR), in 1922 Friedmann found one solution of Einstein
field equation that allows the non-static universe which can
either expand or collapse \cite{fr}. In 1929, there was a great
discovery by Hubble that distant nebulae are in fact other
galaxies outside our Milky Way. By observing these distant
galaxies redshifted, Hubble found that the universe does not stay
static but it is expanding, following his empirical law, dubbed
{\it Hubble's law}, equation (\ref{Hubble}). In 1946, working on
the theory of light-element abundances, Gamow proposed that the
universe at early time should be very hot and dense \cite{Gamow}.
Following Gamow's work, Alpher and Herman \cite{ah} wrote a
programming code predicting that the universe should be filled by
microwave radiation with the black-body spectrum at about $5$ K.
\\

All these ideas lead us to the picture that all galaxies should
start from one infinitely dense and hot point and later expand,
gradually cooling down while expanding. This concept is called
{\bf the hot Big Bang}.

        \subsection{Evidence supporting the hot big bang model}
                  \begin{itemize}
                  \item {\bf Redshift of Distant Galaxies } \\
                  Galaxies' redshift brought about the Hubble 's
                  law and is the first evidence of expanding
                  universe.
                  \item {\bf Primodial Nucleosynthesis} \\
                  The predictions of the abundances of the light
                  elements from nucleosynthesis in the hot Big Bang have been proposed \cite{hoy}
                  and yield results that agree very well with
                  astrophysical observations (for a review see \cite{schramm}).
                  \item {\bf Detection of Cosmic Microwave Background Radiation
                  (CMB)} \\
                  Cosmic microwave background (CMB) was
                  discovered accidentally in 1965 by Penzias and Wilson
                  \cite{pen}. The spectrum of the background is very close to the
                  theoretical prediction since it has
                  black-body (thermal equilibrium) shape and it seems to be very
                  isotropic
                  over the whole sky with the temperature about
                  $2.7$ K. This observed temperature is very close to Alpher and Herman's prediction.
                  Discovery of the CMB provides concrete evidence of the hot Big Bang
                  theory, and against the steady state theory\footnote{The steady state theory is
not discussed here. For more detail, Ref. \cite{na} is
recommended.}.
                 \end{itemize}

                          The Big Bang cosmology relies on the cosmological principle and General Relativity (GR).
                          The universe looks, on large scales, homogenous and
                  isotropic but at smaller scales, it has structure and therefore is neither homogenous nor isotropic.
                  In the early 1990s the results from the COBE mission revealed that there are small fluctuation
                  ($\Delta T/T \sim 1/10^5$) \cite{bennett} in the CMB temperature, and this could be the seeds for origin of
                  structures we see today (see some standard textbooks, e.g. \cite{l&l,coles,kolb} for more discussion).

                  \subsection{Einstein field equation and FLRW metric}
                  The Einstein field equation in GR is
                 \begin{equation}
                 G_{\mu \nu}\,  \equiv \, R_{\mu \nu} - \frac{1}{2}g_{\mu
                 \nu} R \,=\, 8 \pi G T_{\mu \nu} \label{eins1}
                 \end{equation}
                 The field equations can be solved by introducing some assumptions on the metric (or line element) $g_{\mu\nu}$.
                 The metric that satisfies homogeneity and isotropy is the
                 {\bf Friedmann-Lema\^{i}tre-Robertson-Walker metric}
                 (FLRW metric),
                 \begin{equation}
                 ds^2 = dt^2 - a^2(t) \bigg[\frac{dr^2}{1-kr^2} + r^2 d \theta^2 + r^2 \sin^2\theta d \varphi^2 \bigg]
                 \end{equation}

                 where $t$ is proper time, $r$, $\theta$ and $\varphi$
                 are spherical coordinates, $a(t)$ is the scale factor and $k = \pm 1$,$0$ is the spatial curvature mentioned
                 earlier. With the FLRW metric, equation (\ref{eins1}) leads to the the Friedmann and
                 fluid equations.

                 \subsection{Particle horizon}
                 In the FLRW expanding universe, the particle horizon, i.e. the distance that light has travelled from the Big Bang $t=0$
                 until the time $t=t_0$, is

                 \begin{equation}
                 d_H(t_0) \, \equiv \, a(t_0) \bigg[ \int_{t=0}^{t=t_0} \frac{c
                 dt}{a(t)} \bigg]
                 \end{equation}

                 The quantity $\int_{0}^{t_0} \frac{c dt}{a(t)}$ is called the {\it comoving particle horizon}.

                 Since in GR matter curves spacetime, the problems of Newtonian
                 cosmology are solved.

                 \subsection{Hubble time and Hubble length}

                 We can notice from equation (\ref{hubl}) that $H$
                 has a dimension of $t^{-1}$. We call $H^{-1}$
                 {\bf Hubble time}. The Big Bang starts  with
                 $a=0$,
                 so the Hubble time can be used as the
                 approximate age of the universe, as we used it to
                 estimate the age of the dust-dominated universe
                 before. The {\bf Hubble length} at one particular time is the
                 distance that light travels from Big Bang until
                 that time. The Hubble length is defined simply as
                 $cH^{-1}$.

                  \subsection{The Einstein's static universe}
                 In the early 1900s the universe was thought to be static therefore Einstein, trying
                 to match his theory with expectation, introduced
                 the cosmological constant $\Lambda$ into his field equation. This ad hoc
                 fine-tuned value of $\Lambda$ is just enough to hold the universe static.
                 The field equation becomes

                 \begin{equation}
                 R_{\mu \nu} - \frac{1}{2}g_{\mu
                 \nu} R - g_{\mu
                 \nu}\Lambda \,=\, 8 \pi G T_{\mu \nu} \label{eins2}
                 \end{equation}

                 The discovery of the expansion of the universe ruled out the static model of the universe and
                 the cosmological constant fell away.
                 However recently, by observing type Ia supernova
                 \cite{perl}, the expansion of the universe was found to be accelerating and the idea of
                 repulsive gravity e.g. cosmological constant, has come into favour again.
                 \footnote{The cosmological constant could be considered as the vacuum energy according to quantum field
                 theory but there is one big problem: the value of
                 vacuum energy from quantum field theory is about $124$ order greater than the value needed in cosmology!
                 This still remains a problem for cosmologists}

                 \subsection{Cosmological equations}

                 The consistent fluids for the FLRW metric are the perfect fluids mentioned before
                 (the fluids that have no viscosity
                 and no heat conduction in comoving coordinate). We assume that the components
                 of the universe are a mixture of perfect fluids.  The energy momentum tensor for the perfect fluid is
                 \begin{equation}
                 T^{\mu}\:_{\nu} = diag(\rho, -p, -p, -p)
                 \end{equation}
                 The field
                 equation (\ref{eins2}) in the FLRW, perfect fluid-filled universe
                   yields the Friedmann and acceleration
                 equations (\ref{friedmann}) and (\ref{acc}) :

                 \begin{eqnarray}
                 \bigg(\frac{\dot{a}}{a}\bigg)^2 \;&=&\; \frac{8 \pi G \rho}{3} -\frac{k}{a^2} + \frac{\Lambda}{3} \nonumber  \\
                 \frac{\ddot{a}}{a} \;&=&\; - \frac{4 \pi
                 G}{3}\Big( \rho + 3p \Big) + \frac{\Lambda}{3}
                 \nonumber
                 \end{eqnarray}

                 The Einstein field equations imply the
                 energy-momentum conservation equation,
                 \begin{equation}
                 \nabla^{\nu}T^{\mu}\;_{\nu} = 0
                 \end{equation}

                 For a FLRW universe, this gives the fluid equation (\ref{flu})
                 \begin{equation}
                 \dot{\rho}+ 3H\rho(1 + w) = 0  \nonumber
                 \end{equation}

                 \subsection{Density parameter and geometry of the universe}

                 We now rewrite the Friedmann equation (\ref{friedmann})
                 in another form
                 \begin{equation}
                 -k \;=\; a^2 H^2 \Bigg[1 - \frac{8 \pi G}{3H^2}
                 \rho_{tot}
                 \Bigg] \label{friedmann2}
                 \end{equation}
                 where $\rho_{tot}$ is the total energy density of the
                 universe

                 \begin{equation}
                 \rho_{tot} \: = \: \rho_{rad} + \rho_{dust} +
                 \rho_{\Lambda}
                 \end{equation}

                 where $\Lambda$ here includes $\Lambda$ or other dark energy.
                 $\rho_{\Lambda}$ is defined as $\rho_{\Lambda}\equiv \Lambda / 8 \pi G $.

                 According to equation (\ref{friedmann2}), the universe can be flat ($k = 0$) only if
                 \begin{equation}
                 \rho_{tot}  \;=\; \frac{3H^2}{8\pi G}
                 \equiv\rho_{c} \label{friedmann3}
                 \end{equation}
                 This is how we define the critical energy density, $\rho_{c}$, the energy density that is consistent with a flat universe. If $\rho_{tot} > \rho_{c} $,
                 then $k > 0$ and the
                 universe has closed geometry which means that the expansion of the universe could halt at some point
                 and then start to re-collapse.
                 On the other hand if $\rho_{tot} < \rho_{c}$ then $k<0$ and the universe has
                 open geometry. This means that universe will expand
                 forever and in the far future, matter and radiation will be diluted away. Eventually the vacuum energy
                 will dominate the universe and expansion enters the acceleration
                 phase. The sum of
                 the angles of a triangle is less than
                 $180$ degrees for open geometry. The sum is greater
than $180$ degrees for closed geometry and is equal to $180$
degrees for the flat geometry.

                 The Friedmann equation can also be expressed in term
                 of the total density parameter $\Omega_{tot}$ as
                 \begin{equation}
                 1-\Omega_{tot} \;=\;
                 -\frac{k}{a^2 H^2} \label{friedmann4}
                 \end{equation}
                 where
                 \begin{equation}
                 \Omega_{tot} \equiv \rho /
                 \rho_{c} =  \frac{8 \pi G \rho}{3 H^2} \nonumber
                 \end{equation}

                 From equation (\ref{friedmann4}) we can see that $ \Omega_{tot} $
                 determines conditions for spatial curvature of the universe
                 \begin{eqnarray}
                 \Omega_{tot}  &=& 1 \;\, \Rightarrow \;\, k =0 \nonumber \\
                 \Omega_{tot}  &<& 1 \;\, \Rightarrow \;\, k =-1 \nonumber \\
                 \Omega_{tot}  &>& 1 \;\, \Rightarrow \;\, k =1 \nonumber
                 \end{eqnarray}

                 In the year 2000, two CMB detector
                 missions, BOOMERanG and MAXIMA
                 \cite{boom,Maxima} confirmed that the universe's geometry should be very close to flat.
                 The BOOMERang result in 2001 gave $\Omega = 1.02^{+0.05}_{-0.05} $ \cite{boom2}.
                 The WMAP satellite's data released in February 2003 was consistent with this: $\Omega = 1.02^{+0.02}_{-0.02}$ \cite{wmap1}.
                 \\

                 \subsection{Cooling down with expansion}

                 The temperature of the present universe is about 2.728 K.
                 The background radiation of the universe has a black-body spectrum,
                 therefore the relation between radiation energy density $\varepsilon $
                 and temperature $T$ of the universe is
                 \begin{equation}
                 \varepsilon\; = \;\rho_{rad} c^2  \;=\; \alpha T^4
                 \end{equation}
                 where $\alpha \equiv \pi^2{k_B}^4 / 15 \hbar^3
                 c^3$, $k_B$ is Boltzmann's constant and $\hbar$ is the reduced
                 Planck constant. Equation (\ref{rhorad}) for radiation implies that
                 $\rho =  \rho_{0}
                 \Big(\frac{a}{a_0}\Big)^{-4}$. This leads to the
                 relation
                 \begin{equation}
                 T \; \propto \; \frac{1}{a}
                 \end{equation}
                 and by equation (\ref{scale}) we have
                 \begin{equation}
                 T \; \propto \; \frac{1}{t^{2/[3(1+w)]}}
                 \end{equation}

                 This results imply that the universe with dust or
                 radiation domination {\it should be hotter and smaller
                 at earlier time}. This is the main concept of the
                 standard hot Big Bang cosmology.

\subsection{Problems of the hot big bang model}
                 Despite the fact that the hot Big Bang theory can explain the redshift,
                 abundance of primordial nucleus of light elements and the existence of the CMB,
                 there are still some puzzles that can not be explained in the
                 hot Big Bang framework. These problems are as
                 follows.

            \subsubsection{Flatness problem}
            Using the Friedmann
            equation (\ref{friedmann4}) to compare the value of $|\Omega_{tot}(t)-1|$
            in
            the early universe and today, we can show that
            \begin{equation}
            |1-\Omega_{tot}(t)| \: \propto \: a^2 \: \propto \: t
            \label{o1}
            \end{equation}
            during the radiation-domination era and
             \begin{equation}
            |1-\Omega_{tot}(t)| \: \propto \: a \: \propto \: t^{2/3}
            \label{o2}
            \end{equation}
            in the dust-domination era.
            These equations tell us how the density parameter
            evolves with time.
            \\

These equations show that if $\Omega_{tot}$ is not exactly $1$,
then its difference from $1$ grows with expansion.
 The universe observed today is very flat, with
$|1-\Omega_{tot,0}| \sim
            0.05$, therefore the universe at early time should be
            very very close to flat. Working out the equation (\ref{o1}) and (\ref{o2}),
            we would require the universe to be extremely flat e.g. $|\Omega_{tot}-1| \sim 10^{-16}$ at
            nucleosynthesis. If the initial value of
            $\Omega_{tot}$ was not very close to $1$, the universe would either re-collapse or
            expand and dilute very quickly and will not evolve to our universe today. The hot Big Bang theory is
            not able to explain how and why the universe was almost
            perfectly flat in the first place.

            \subsubsection{Horizon problem}
            The CMB across the sky looks roughly isotropic. However there is something
            unacceptable about
            this fact. The CMB was emitted when the universe was
            about 300,000 years old at recombination and the horizon at that time
            was about 300,000 light years which is
            approximately one degree in the sky today. Any area in universe that has size greater
            than 1 degree therefore should not be able to be thermalized by photons and hence
            the smooth-looking thermal-equilibrium CMB sky should not be seen
            today. This can not be explained by the hot Big Bang
            picture.

            \subsubsection{Magnetic monopole problem}
            The theory of particle physics has predicted many of
            the exotic particles that could be created in the
            early universe. Some examples are magnetic monopoles, gravitinos, moduli fields and other
            higher dimensional objects from topological defects
            e.g. cosmic strings, domain walls and textures. Where
            are they today?.

            \subsubsection{Origin of structure problem}
           The CMB anisotropies observed first by COBE
             in 1992 can not be explained by the hot Big Bang
            picture \cite{sato}. The scale of the anisotropies is too large to
            be produced during the time from the Big Bang to the time of
            decoupling, because it should be thermalized according to the Big Bang
picture.
            This means that anisotropies must have been a part the initial condition which is very unlikely.
            Since the anisotropies in CMB are signatures telling us
            how structures in the universe were formed, the hot Big Bang theory fails to provide an answer
            for a theory of structure formation.

    \section{Inflationary Universe}
            Inflation proposed by Guth, Sato, Albrecht, Steinhardt and Linde in 1980s is the period of the early universe that undergoes
            an accelerating phase \cite{guth}. Inflation of the universe is equivalent to
            \begin{equation}
            \ddot{a}  \: > \: 0
            \end{equation}
            so that
            $ \dot{a}$ increases during the inflation phase. As a
            result the comoving Hubble length $(aH)^{-1} $ must be decreasing in this
            phase, i.e.
            \begin{equation}
            \frac{d}{dt} \Big(\frac{1}{aH}\Big)  \: < \:0
            \end{equation}
            The acceleration equation (\ref{acc})
            requires that
            \begin{equation}
            \rho + 3p \,<\, 0 \;\; \Rightarrow \;\; p < -\frac{\rho}{3}
            \label{np}
            \end{equation}
            Inflation is able to solve the problems of the hot
            Big Bang as will be explained. It does not substitute the hot Big Bang idea
            but instead it adds on some ideas and also modifies the
            hot Big Bang model.

            \subsection{Solving the hot big bang problems}
            \begin{itemize}
            \item {\bf Flatness Problem}
            The flatness problem can be solved directly with
            inflation. In equation (\ref{friedmann4}), to make the difference $\Omega_{tot} -
            1$ smaller, $aH$ must increase. The accelerating
            expansion yields directly an increase in $aH (=\dot{a})$. If enough increment in $aH$
            has been made by inflation to drive
$\Omega_{tot}$ very close to  $1$,  we can explain why the final
density parameter value is very close to one.

            \item {\bf Horizon Problem}
            To obtain a universe that looks almost isotropic
            today, we need a universe that has causal contact
            over the whole sky before inflation. In this scenario, during inflation any small region of the
            universe will inflate to a very large region. This can explain why today sky looks almost
            the same in all directions.
            \item {\bf Monopole Problem}
            The super-fast expansion of inflation can dilute away these
            relics predicted by particle physics. That is why we do not detect them today.

            \item {\bf Origin of Structure Problem}
            The scalar field that drives inflation (see below)
            experiences quantum fluctuations. These fluctuations are stretched by inflation to scales where they
            cause ripples in the CMB temperature. These ripples
            are the seeds that lead to the formation of structures
            and galaxies.
             \end{itemize}

            \subsection{What drives inflation?}
To produce inflation, some matter that can yield this acceleration
phase has to be dominant at that period. The candidate from
particle physics responsible for driving inflation is a scalar
field, which is called the {\it inflaton field}, $\phi(t)$. The
pressure and energy of the inflaton field are given by
\begin{eqnarray}
p_{\phi} &=& \frac{1}{2}\dot{\phi}^2 - V(\phi)
\\ \rho_{\phi} &=&\frac{1}{2}\dot{\phi}^2 + V(\phi)
\end{eqnarray}

where $V(\phi)$ is the potential energy. During the inflation
phase the inflaton field dominates the universe. The scalar
field-dominated Friedmann equation reads
\begin{equation}
H^2 = \frac{ 8 \pi G}{3}\bigg[\frac{1}{2}\dot{\phi}^2 +
V(\phi)\bigg]
\end{equation}
If we use the scalar field pressure and energy density in the
fluid equation, we obtain the Klein-Gordon equation
\begin{equation}
\ddot{\phi} + 3H\dot{\phi} = - V'(\phi)
\end{equation}
where $'$ denotes differentiation with respect to $\phi$. This
equation governs the dynamics and energy conservation of the
inflaton field. The condition for acceleration from equation
(\ref{np}) requires that
\begin{equation}
\dot{\phi}^2 < V
\end{equation}
and it implies that inflation can be sustained when the field
moves very slowly, i.e. the kinetic energy  $\dot{\phi}^2$ is very
small. This is called the {\bf slow-roll approximation}. We can
use this approximation as a criterion for inflation to happen.
With very small $\dot{\phi}^2$, the Friedmann equation is
approximately
\begin{equation}
H^2 \simeq \frac{8 \pi G}{3}V
\end{equation}
and the Klein-Gordon equation gives
\begin{equation}
\dot{\phi} \simeq - \frac{V'}{3H}
\end{equation}
Since $|V'|$ is very small, $H$ is nearly constant, so that $a(t)$
grows nearly exponentially ($\frac{\dot{a}}{a} \simeq$ constant).
The amount of inflation can be measured in term of the e-folding
number, $N$ given by equation (\ref{N}). Normally at least $50$
e-foldings are needed to solve the hot Big Bang problems. Refs.
\cite{l&l,riotto1} are recommended for further reading on
inflation.

\subsection{After inflation}

\subsubsection{Reheating}
During inflation the universe is supercooled by very rapid
expansion. Inflation stops when the field begins to roll faster
down the potential and the slow-roll approximation breaks down.
After inflation, the inflaton field begins to decay, producing
matter and radiation. The radiation produced by this decaying
starts to {\it reheat} the universe, providing the standard hot
Big Bang phase.

\subsubsection{Structure formation}
Predicting the seeds for structure formation is the true merit of
inflationary theory. Inflation creates perturbations which are in
the form of density (scalar) perturbations and gravitational
waves. Density perturbations are created from quantum vacuum
fluctuations of the inflaton field, which leave an imprint as
inhomogeneities in the radiation and matter at reheating
\cite{haw}. The wavelength of the fluctuations is stretched by
inflation. The fluctuations therefore become bigger in size and
then provide seeds of gravitational instability for structure to
form. However, another type of perturbation, the gravitational
wave, does not contribute to structure formation. These
perturbations instead cause ripples in the geometry of space-time.
More detail of the structure formation, Ref. \cite{coles} is
recommended.
\\

Observational cosmologists have information on the present
structures in the universe, and they try to match cosmic structure
patterns with the information found in the CMB
 anisotropies which were created in the early universe.

\section{Conclusion}
I have outlined the historical background and derived the
cosmological equations using Newtonian cosmology. Basic ideas of
the hot big bang and relativistic cosmology are also given here.
Finally I have introduced briefly inflation and explained how
structure forms in an elementary language.
\\

{\flushleft{ \large \bf Acknowledgement:}} I would like to thank
organizations which contributed to my stay in Britain during
1998-2003: the Royal Thai Government Scholarship, the British-Thai
Scholarship Scheme (BTSS), the Institute of Cosmology and
Gravitation (ICG) at University of Portsmouth and Department of
Physics of Naresuan University. I also thank Roy Maartens and Mark
Hindmarsh who are my supervisors and who taught me so much.
Special thanks is given to Roy Maartens for editing my lecture
notes. I am grateful to those who have been through hard time in
organizing this school with me: Bruce Bassett, Rachan Rangdee,
Sukruedee Nathakaranakule, Thiranee Khumlumlert, Anakaorn
Suwattiphun and Kiattisak Tepsuriya. I also thank David Wands for
 useful discussions. I really appreciate Cindy Lau for giving me
days and nights to spend on work. Finally I thank the Tah Poe
Group of Theoretical Physics (Tptp) and its young-people idealism
that keeps us collaborating and working together.

\setlength{\baselineskip}{12pt}
\renewcommand{\baselinestretch}{1.8}


\begin{thebibliography}{99}

\bibitem{Harrison} E. R. Harrison, {\sl Cosmology: The Science of the Universe (2nd
Ed.)}, Cambridge University Press, Cambridge (2000).
\bibitem{burin} B. Gumjudpai in {\sl Proceedings from the First Tah Poe School on Cosmology
2002} (with B. Bassett, C. Dariescu and M. Dariescu), Department
of Physics, Naresuan University, Phitsanulok (2002).
\bibitem{bennett} C. L. Bennett, {\it Astrophys. J.} {\bf
464}, L1-4 (1996) [{\sf astro-ph/9601067}].

\bibitem{fr} A. Friedmann, {\it Z. Phys.} {\bf 10}, 377 (1922).

\bibitem{mc} W.H. McCrea and E. Milne, {\it Quarterly J. of Math.} {\bf
5} 73 (1934).

\bibitem{Hub} E. Hubble, {\it Proc. Nat. Acad. Sci.} {\bf 15}, 168
(1929).
\bibitem{Hub1} E. Hubble, {\sl The Realm of the Nebulae}, Yale University Press, New Haven, CT (1936).
\bibitem{Hub2} E. Hubble, {\sl The Observational Approach to
Cosmology},
 Oxford University Press, Clarendon Press, Oxford
(1937).

\bibitem{aegean1} A. Paraskevopoulos in
{\sl Cosmological Crossroads: An Advanced Course in Mathematical,
Physical and String Cosmology}, S. Cotsakis and E. Papantonopoulos
(Editors) Springer, Berlin (2002) {\footnotesize(The Proceedings
from the First Aegean Summer School on Cosmology, Samos Island,
Greece 2001)}.

\bibitem{liddle}  A. R. Liddle,  {\sl An Introduction to Modern Cosmology}, Wiley, Chichester (1998).

\bibitem{perl} S. Perlmutter, M. Della Valle et al., {\it Nature} {\bf
391}, 51 (1998) [{\sf astro-ph/9712212}].
\\ S. Perlmutter, G. Aldering et al.,  {\it Astrophys. J.}  {\bf
517}, 565 (1999) [{\sf astro-ph/9812133}].
\\ P. M. Garnavich et al., {\it Astrophys. J.}  {\bf
493}, 53 (1998) [{\sf astro-ph/9710123}].
\\ A. G. Riess et al., {\it Astronomical J.}  {\bf
116}, 1009 (1998) [{\sf astro-ph/9805201}].

\bibitem{sth} P. J. Steinhardt's webpage \\ ({\tt
http://feynman.princeton.edu/$\sim$steinh/royal.pdf}).


\bibitem{boom} P. de Bernardis et al., {\it Nature}
{\bf 404}, 955 (2000) [{\sf astro-ph/0004404}].

\bibitem{Maxima} S. Hanany et al., {\it Astrophys. J.} {\bf 545}, L5 (2000) [{\sf astro-ph/0005123}]. \\
                 A. Balbi et al., {\it Astrophys. J.} {\bf 545}, L1-L4 (2000) [{\sf astro-ph/0005124}].

\bibitem{boom2} P. de Bernardis et al., {\it Astrophys. J.}
{\bf 564}, 559 (2002) [{\sf astro-ph/0105296}].

\bibitem{wmap1} C. L. Bennett et al. (2003) [{\sf astro-ph/0302207}].

\bibitem{wmaph} D. N. Spergel et al. (2003) [{\sf astro-ph/0302209}]. \\ WMAP
Collaboration 2003 ({\tt http://map.gsfc.nasa.gov}).


\bibitem{Gamow} G. Gamow, {\it Phys. Rev.} {\bf 70}, 572 (1946).
\\ R. A. Alpher, H. Bethe and G. Gamow, {\it Phys. Rev.} {\bf 73}, 803
(1948). \\  G. Gamow, {\it Phys. Rev.} {\bf 74}, 505 (1948).

\bibitem{ah} R. A. Alpher and R. Herman, {\it Nature} {\bf 162},
774 (1948).

\bibitem{hoy} F. Hoyle and R. J. Tayler, {\it Nature} {\bf 203},
1108 (1964). \\ P. J. E. Peebles, {\it Astrophys. J.} {\bf 146},
542 (1966). \\ R. V.Wagoner, W. A. Fowler and F. Hoyle, {\it
Astrophys. J.} {\bf 148}, 3 (1967).

\bibitem{schramm} D. N. Schramm and R .V. Wagoner, {\it Ann. Rev. Nucl. Part. Sci.} {\bf 27}, 37
(1979).\\ S. M. Austin, {\it Prog. Part. Nucl. Phys.} {\bf 7}, 1
(1981).\\ D. N. Schramm, {\sl The Big Bang and Other Explosions in
Nuclear and Particle Astrophysics}, World Scientific (1995).

\bibitem{pen} A.A.Penzias and R.W. Wilson, {\it Astrophys. J.} {\bf
142}, 419 (1965).



\bibitem{l&l} A. R. Liddle and D. Lyth,
{\sl Cosmological Inflation and Large-Scale Structure}, Cambridge
University Press, Cambridge (2000) \\ A. R. Liddle and D. Lyth,
{\it Phys. Report} {\bf 231}, 1 (1993) [{\sf astro-ph/9303019}].

\bibitem{riotto1} A. Riotto (2002) [{\sf hep-th/0210162}].

\bibitem{coles} P. Coles and F. Lucchin, {\sl  Cosmology: The Origin and Evolution of Cosmic Structure}, Wiley (2002).
\bibitem{kolb} E. W. Kolb and M. S. Turner, {\sl  The Early Universe}, Addison-Wesley (1990).


\bibitem{na}  F. Hoyle, G. Burbidge and Jayant V. Narlikar
{\sl A Different Approach to Cosmology : From a Static Universe
through the Big Bang towards Reality}, Cambridge University Press,
Cambridge (2000).

\bibitem{guth} A. H. Guth, {\it Phys.Rev. D} {\bf 23}, 347 (1981). \\
               K. Sato, {\it Mon. Not. R. Astro. Soc.} {\bf 195}, 467
               (1981).\\
                 A. Albrecht and P. J. Steinhardt, {\it Phys. Rev. Lett.} {\bf 48}, 1220
                 (1982). \\
                 A. D. Linde, {\it Phys. Lett.} {\bf 108B}, 389
                 (1982).
\bibitem{sato} Y. Suto, K. Sato and H. Kodama, {\it Astrophys. J. Lett.} {\bf 292}, L1 (1985).
\bibitem{haw} A. H. Guth and S. Y. Pi,  {\it Phys. Rev. Lett.} {\bf 49}, 1110
(1982).
\\ S. W. Hawking, {\it Phys. Lett. B} {\bf 115}, 295 (1982).
\\ A. A. Starobinsky, {\it Phys. Lett. B} {\bf
117}, 175 (1982).
\end{thebibliography}
\end{document}